\documentclass[twocolumn,showpacs,preprintnumbers,amsmath,amssymb, prb]{revtex4}
\usepackage{graphicx}
\usepackage{dcolumn}
\usepackage{bm}
\usepackage{multirow}
\usepackage{hyperref}

\begin{document}
\preprint{APS/123-QED}
\title{Anisotropic magnetic properties and giant magnetocaloric effect of PrSi single crystal}
\author{Pranab Kumar Das}
\affiliation{Department of Condensed Matter Physics and Materials Science, Tata Institute of Fundamental Research, Homi Bhabha Road, Colaba, Mumbai 400 005, India.}


\author{Amitava Bhattacharyya}
\altaffiliation[Present address:]{ISIS Facility, Rutherford Appleton Laboratory, Chilton, Didcot Oxon, OX11 0QX, UK}
\affiliation{Department of Condensed Matter Physics and Materials Science, Tata Institute of Fundamental Research, Homi Bhabha Road, Colaba, Mumbai 400 005, India.}

\author{Ruta Kulkarni}
\affiliation{Department of Condensed Matter Physics and Materials Science, Tata Institute of Fundamental Research, Homi Bhabha Road, Colaba, Mumbai 400 005, India.}
\author{S. K. Dhar}
\affiliation{Department of Condensed Matter Physics and Materials Science, Tata Institute of Fundamental Research, Homi Bhabha Road, Colaba, Mumbai 400 005, India.}

\author{A. Thamizhavel}
\affiliation{Department of Condensed Matter Physics and Materials Science, Tata Institute of Fundamental Research, Homi Bhabha Road, Colaba, Mumbai 400 005, India.}

\date{\today}

\begin{abstract}
Single crystal of PrSi was grown by Czochralski method in a tetra-arc furnace. Powder x-ray diffraction of the as grown  crystal revealed that PrSi crystallizes in FeB$-$type structure with space group \textit{Pnma} (no. 62). The anisotropic magnetic properties were investigated by means of magnetic susceptibility, isothermal magnetization, electrical transport and heat capacity measurements. Magnetic susceptibility data clearly indicate the ferromagnetic transition in PrSi with a $T_{\rm C}$ of 52~K.  The relative easy axis of magnetization was found to be the [010] direction. Heat capacity data confirm the bulk nature of the transition at 52~K and exhibit a huge anomaly at the transition. A sharp rise in the low temperature heat capacity  has been observed  (below 5~K) which is attributed to the $^{141}$Pr nuclear Schottky heat capacity arising from the hyperfine field of the Pr moment.  The estimated Pr magnetic moment 2.88~$\mu_{\rm B}$/Pr from the hyperfine splitting is in agreement with the saturation magnetization value obtained from the magnetization data measured at 2~K. From the crystal electric field (CEF) analysis of the magnetic susceptibility, magnetization and the heat capacity data it is found that the degenerate $J~=~4$ Hund's rule derived state of Pr$^{3+}$-ion splits into nine singlets with an overall splitting  of 284~K, the first excited singlet state separated by just 9~K from the ground state. The magnetic ordering in PrGe appears to be due to the exchange generated admixture of low lying crystal field levels.  Magnetocaloric effect (MCE) has been investigated from magnetization  data along all the three principal crystallographic directions. Large magnetic entropy change, $-\Delta S_M = $22.2~J/kg~K, and the relative cooling power, RCP = $460$~J/kg, characteristic of giant magneto caloric effect  are achieved near the transition temperature ($T_{\rm C}$~=~52~K) for $H =$~70 kOe  along $[010]$. Furthermore, the PrSi single crystal exhibits a giant MCE anisotropy.
\end{abstract}

\maketitle

\section{Introduction}
The RX (R = rare earth, and X = Si or Ge) compounds crystallize in two closely related orthorhombic crystal structures, namely CrB and FeB type structure, depending upon R and heat treatment during crystal growth. PrGe and some RSi (R = Tb, Dy, Ho) compounds show dimorphism by crystallizing in both CrB and FeB type structures.  Most of the RX compounds order antiferromagnetically at low temperature, while PrX and NdX (X = Si or Ge) have ferromagnetic ground state. In our previous study on CeGe and PrGe single crystals~\cite{Pranab_CeGe, Pranab_PrGe}, we have reported interesting magnetic properties of these two compounds. CeGe orders antiferromagnetically below 10.5~K, and a superzone gap opens along all the three principal directions of the orthorhombic unit cell, inferred from the upturn of the resistivity at $T_{\rm N}$, in contrast to rare-earth elements where the gap opens up only along the hexagonal $c$-axis~\cite{Mackintosh}.  On the other hand PrGe orders antiferromagnetically at $T_{\rm N}$ = 44~K, which is followed by a ferromagnetic transition at 41.5~K. In an earlier report~\cite{Nguyen} on PrSi and NdSi, Nguyen et. al., have studied the magnetic structure of PrSi and NdSi polycrystalline samples by neutron diffraction. They have concluded that both compounds have non-collinear ferromagnetic structure with a small antiferromagnetic component, and the moments lie in the $(a-c)$ plane.  Furthermore, the Nguyen et al., mentioned that the non-collinear ferromagnetic structure in PrSi can only be accounted for by a large anisotropic exchange interaction and a strong crystal field anisotropy. However, in a recent study on polycrystalline PrSi, Snyman and Strydom~\cite{Strydom},  reported that the ferromagnetic moments lie parallel to the crystallographic $b$-axis. In order to remove the discrepancy between these two reports we felt an investigation based on a single crystal of PrSi would be illuminating.  Towards that end, we have grown a single crystal of  PrSi and we have measured its electrical resistivity, magnetization and heat capacity. A detailed crystal field analysis of the magnetization and heat capacity has been carried out.  We also present our results pertaining to the anisotropic magnetocaloric effect in PrSi.

\section{Experiment}
The binary phase diagram of Pr and Si reported by Massalski \textit{et al.}~\cite{Massalski}, shows that PrSi  melts congruently at 1657~$^\circ$C thus indicating that PrSi can be grown directly from the stoichiometric melt.  A single crystal of PrSi was grown by Czochralski crystal pulling method using a tetra-arc furnace (Technosearch Corporation, Japan) under argon atmosphere. High purity starting elements of Pr (99.9\%) and Si (99.999\%) were taken  in the stoichiometric ratio $1:1$, and melted in the tetra-arc furnace several times to ensure homogeneity.  A tungsten rod was used as the seed crystal and after optimising the initial conditions of the growth, the crystal was pulled at a constant rate of $10$~mm/h. The grown crystal had a length of about $7$~cm, and it had a cleavage plane (010) perpendicular to the growth direction.  A polycrystalline sample of LaSi was prepared by arc melting to serve as a non-magnetic reference. Powder x-ray diffraction (XRD) was performed using a PANalytical X-ray diffractometer with monochromatic Cu K$\alpha$ radiation and the crystallographic orientation of the grown crystal was done by back reflection Laue diffraction technique with a polychromatic source of X-ray beam.  The crystal was  cut in a bar shape using a spark-erosion cutting machine for the anisotropic physical property measurements.    Magnetization and electrical transport measurements were performed on well oriented single crystals using Quantum Design Vibration Sample Magnetometer (VSM), SQUID magnetometer and home made electrical resistivity set-up, respectively. Heat capacity was measured in the temperature interval from 1.8~K to 300~K using a Quantum Design Physical Property Measurement System (PPMS). 

\section{Results and discussion}
\subsection{X-ray diffraction}
\begin{figure}[!]
\includegraphics[width=0.45\textwidth]{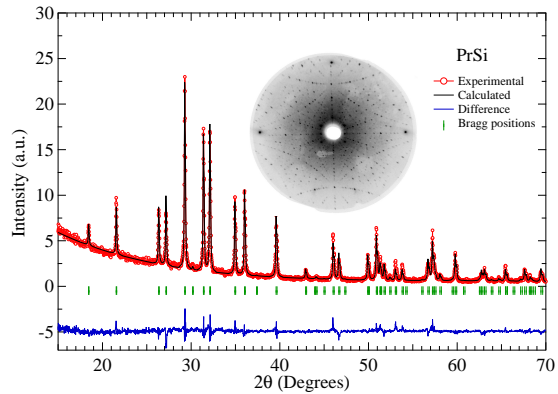}
\caption{\label{Fig1}(Color online) Powder x-ray diffraction pattern of PrSi along with the Rietveld refinement. A representative Laue pattern, corresponding to (010) plane is shown in the inset.}
\end{figure}

PrSi crystallizes in FeB$-$type orthorhombic crystal structure with the space group \textit{Pnma} (no. 62). Pr and Si atoms occupy the $4c$ crystallographic site~\cite{Nguyen}, which possesses a monoclinic site symmetry. In order to confirm the phase purity and to determine the lattice constants of the grown crystal, we recorded the powder x-ray diffraction pattern on finely ground sample. A Rietveld analysis performed on the powder pattern using FullProf software package~\cite{Fullprof} is shown in Fig.~\ref{Fig1}. The lattice constants obtained from the Rietveld analysis, $a = 8.240(2)$~\AA, $b = 3.942(1)$~\AA~ and $c = 5.921(1)$~\AA, respectively, are in good agreement with the previous report~\cite{Nguyen}. The  positional parameters of Pr and Si atoms that occupy the $4c$ site are (0.179, 0.25, 0.612) and (0.040, 0.25, 0.120), respectively.  Furthermore, the composition of the grown crystal was also confirmed using Energy Dispersive Analysis by X-ray (EDAX). The single crystal was oriented along the three principal crystallographic directions by means of back reflection Laue diffraction method. We observed well defined Laue pattern for all three principal crystallographic directions ascertaining the good quality of the grown crystal.  The symmetry pattern of the Laue diffraction also confirmed the \textit{Pnma} space group of PrSi. A representative Laue pattern corresponding to the (010) plane is shown in the inset of Fig.~\ref{Fig1}. 

\subsection{Electrical Resistivity}

\begin{figure}[!]
\includegraphics[width=0.45\textwidth]{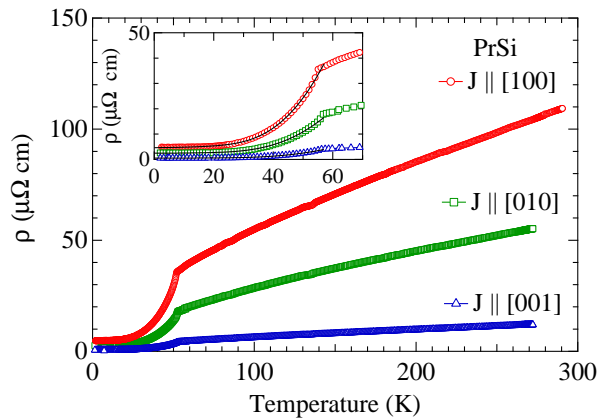}
\caption{\label{Fig2}(Color online) Temperature dependence of electrical resistivity of PrSi.  The inset shows the low temperature part of the electrical resistivity.  The solid lines are the fits to spin-wave gap model.}
\end{figure}

The temperature dependence of the electrical resistivity of PrSi measured along the three principal crystallographic directions [100], [010] and [001], respectively is shown in Fig.~\ref{Fig2}.  Close to room temperature the resistivity values are  $\rho_{\rm [100]} = 109~\mu \Omega$cm, $\rho_{\rm [010]} = 55~\mu \Omega$cm and $\rho_{\rm [001]} = 12~\mu \Omega$cm, thus indicating a large anisotropy reflecting the orthorhombic crystal structure.  The electrical resistivity decreases as the temperature is decreased, with a very broad curvature typical of the rare-earth systems.  The broad curvature in the electrical resistivity is mainly attributed to the thermal de-population of the crystal field levels as the temperature is reduced.  The electrical resistivity falls off rapidly below 52~K, due to the reduction in the spin-disorder scattering, where the system undergoes a magnetic ordering.  It is evident from the magnetization data to be discussed later, PrSi 
undergoes a ferromagnetic ordering at $T_{\rm C} = 52$~K. The low temperature electrical resistivity can be described by the spin-wave gap model.  In the case of non-cubic (anisotropic) ferromagnets,  the dispersion relation of the spin-wave spectrum is given by~\cite{Andersen, Larrea} $\hbar \omega_{\rm m} = \Delta + Dq^2_{\rm m}$, where $\omega_{\rm m}$ and $q_{\rm m}$ are the frequency and wave vector of magnons, respectively and $D$ is the spin-wave stiffness and $\Delta$ is the gap in the spin-wave spectrum.  The gapped magnons give rise to a temperature dependent resistivity for $k_{\rm B}T~\ll~\Delta$, which takes the form,

\begin{equation}
\label{eqn1}
\rho_(T) = \rho_0 + a T^2 + b T \Delta \left(1 + \frac{2T}{\Delta}\right) {\rm exp} {\left(\frac{-\Delta}{T}\right)}.
\end{equation}

The second term in the above expression is the usual Fermi liquid term and the third term is the contribution from  magnons. In Eqn.~\ref{eqn1}, $\rho_0$ is the residual resistivity,  the coefficient  $a$ determines the degree of electron-electron scattering and $b$ is a constant for the given material and depends on the spin wave stiffness $D$. The low temperature resistivity fits very well to the above expression as shown in the inset of Fig.~\ref{Fig2}.  The parameters obtained from the fit are given in Table~\ref{Table1}.
\begin{table}
\begin{center}
\begin{tabular}{ccccc}
\hline 
 & $\rho_{0}$ & a & b & $\Delta$
\tabularnewline
 & ($\mu \Omega$~cm) & ($\mu \Omega$~cm/K$^2$)  & ($\mu \Omega$~cm/K$^2$) & (K)\\ 
\hline \\
$\rho_{\rm [100]}$ & 4.721  & $1.91 \times 10^{-3}$ & $2.88 \times 10^{-2}$  & 128 \\ \\
$\rho_{\rm [010]}$ & 2.477 & $9.188 \times 10^{-4}$ & $9.53 \times 10^{-3}$ & 124 \\ \\
$\rho_{\rm [001]}$ & 0.525 & $2.242 \times 10^{-4}$  & $2.58 \times 10^{-3}$ & 132 \\ \\
\hline
\end{tabular}
\caption{\label{Table1} Fitting parameters of the resistivity data along the three principal directions described in Eqn.~\ref{eqn1}.}
\end{center}
\end{table}
The residual resistivity ratio (RRR) amounts to 23, 22, and 23 for current parallel to [100], [010], and [001], respectively. The relatively small values of residual resistivity and the high value of RRR indicate the good quality of the single crystal.

\subsection{Magnetization and crystal field analysis}

\begin{figure}[!]
\includegraphics[width=0.4\textwidth]{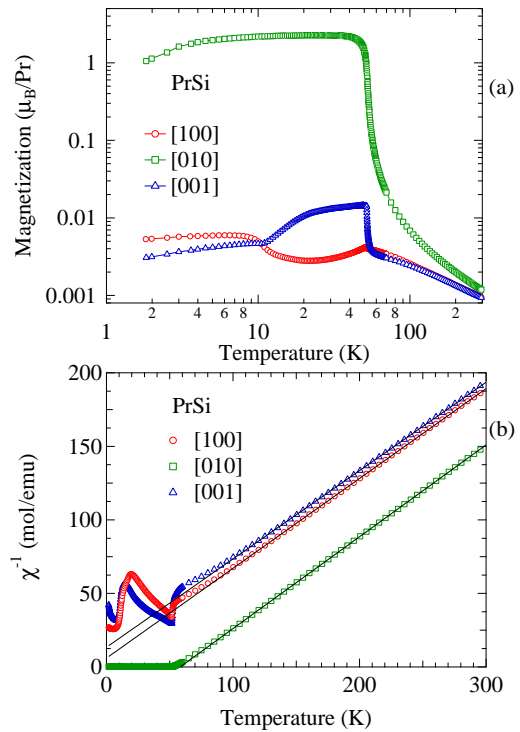}
\caption{\label{Fig3}(Color online) (a) Temperature dependence of magnetization of PrSi along [100], [001] and [010] in an applied field of 1~kOe plotted on a log-log scale, (b) shows the reciprocal magnetic susceptibility of PrSi.  The solid lines represent the Curie-Weiss fit as mentioned in the text.  }
\end{figure}

The magnetization versus temperature measured along the three principal crystallographic directions in the temperature range 1.8 to 300~K in an applied magnetic field of 1000~Oe is shown in Fig~\ref{Fig3}(a).   The magnetization along the [010] direction increases sharply at 52~K signalling the ferromagnetic nature of the magnetic ordering.  Similarly, for $H~\parallel$~[100] and [001] directions, the magnetization shows an anomaly at 52~K; however, relative to [010] the magnetic behavior along these two directions is relatively complex.  It is evident from the figure that the anisotropy in the magnetization is relatively smaller for $H~\parallel$~[100] and [010] respectively, while the magnetization is quite large along the [010] direction, thus indicating [010] as the easy axis of magnetization. 
The reciprocal magnetic susceptibility is shown in Fig.~\ref{Fig3}(b).  For temperatures above 100~K, the magnetic susceptibility follows the Curie-Weiss law $\chi (T) = \mu_{\rm eff}^2/(8 (T-\theta_{\rm p}))$, where $\mu_{\rm eff}$ is effective magnetic moment and $\theta_{\rm p}$ is the paramagnetic Curie temperature.  From the fit of Curie-Weiss expression to the data in the temperature range 100 to 300~K, $\mu_{\rm eff}$ and  $\theta_{\rm p}$ values were estimated to be $3.62(2)~\mu_{\rm B}/$Pr and $-10.3(2)$~K; $3.58(1)~\mu_{\rm B}/$Pr and $58.2(1)$~K; $3.654(3)~\mu_{\rm B}/$Pr and $-22.9(3)$~K for $H~\parallel$~[100], [010], [001], respectively. The experimental value of the effective moment $\mu_{eff}$ is close to the theoretical free ion value of $Pr^{3+}$ based on the Russell-Saunders $L-S$ coupling $(g_{\rm J} \sqrt{J (J+1)} = 3.58)$, where $g_{\rm J}=4/5$ and $J=4$ are the Lande's factor and the total angular momentum for the $4f^2$ configuration of Pr.   The Curie-Weiss temperature is negative for [100] and [001] crystallographic directions, while it is positive along the [010] direction. However, the polycrystalline average is positive (8.3~K), which is in conformity with the ferromagnetic nature of ground state.  It may be noted that $\theta_{\rm p}$ along [010] direction is comparable to $T_{\rm C} = 52$~K.

We analysed the magnetic susceptibility of PrSi based on the crystalline electric field (CEF) model, to understand the magnetocrystalline anisotropy and to estimate the crystal field level splitting. The Pr atom in PrSi occupies the $4c$ Wyckoff's position which has the monoclinic site symmetry. Hence the free ion Pr$^{3+}$ 9-fold degenerate level will split into 9 singlets.   In spite of the singlet ground state PrSi undergoes a ferromagnetic transition at $T_{\rm C}=52$~K.  For a singlet level ground state system,  there is a threshold value of the magnetic exchange interaction between the ground state and the first excited state.  Below this threshold value the system is non-magnetic and when the magnetic exchange interaction is strong enough the system transforms into a magnetically ordered state.  Usually, this happens when the energy separation between the ground state and the first excited state is sufficiently low. 

Although the Pr atom possesses monoclinic site symmetry, in order to reduce the number of fitting parameters we have used the CEF Hamiltonian for the orthorhombic site symmetry, 

\begin{equation}
\begin{split}
\label{eqn2}
\mathcal{H}_{\rm CEF} = B_2^0 O_2^0 + B_2^2 O_2^2 + B_4^0 O_4^0 + B_4^2 O_4^2 + B_4^4 O_4^4 \\ + B_6^0 O_6^0 + B_6^2 O_6^2 + B_6^4 O_6^4 + B_6^6 O_6^6, 
\end{split}
\end{equation}

where $B_l^m$ and $O_l^m$ are the crystal field parameters and the Steven's operators~\cite{Hutchings, Stevens}, respectively.  Here we have defined the quantization axis or the $z-$axis as the [010] direction, [100] direction as the $y-$axis and the [001] direction as the $x-$axis.  

\begin{figure}[!]
\includegraphics[width=0.4\textwidth]{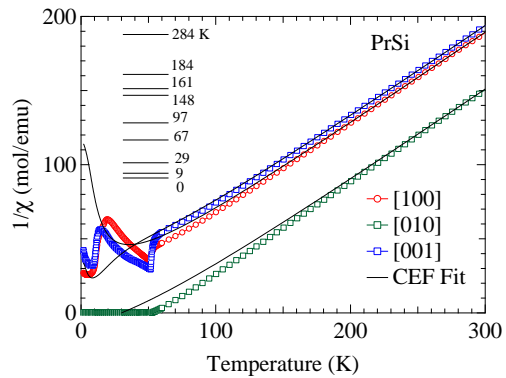}
\caption{\label{Fig4}(Color online) Inverse $\chi$ vs. $T$ plot, the solid lines show the fit based on the crystalline electric field model (see text for details). The crystal field energy levels are shown }
\end{figure}

The magnetic susceptibility $\chi_{\rm i}$ calculated from the CEF Hamiltonian and including the molecular field contribution is given by
\begin{equation}
\label{eqn3}
\frac{1}{\chi_i} = \left(\frac{1}{\chi_{{\rm CEF}i}}\right)-\lambda_{i},  ~~~~(i = x, y, z)
\end{equation}

where $\lambda_{i}$ is the molecular exchange field constant and the expression for $\chi_{{\rm CEF}i}$ is defined in Ref~\onlinecite{Pranab_CeMg3}. The reciprocal magnetic susceptibility along the three principal directions has been calculated based on the above CEF model and the crystal field parameters and the molecular field constants are obtained.   The calculated reciprocal susceptibility and the energy levels are shown in Fig.~\ref{Fig4}. It is evident from the figure that the calculated CEF susceptibility reasonably matches with the experimental data despite the approximation that we have made above. The crystal field parameters and the obtained energy levels are listed in Table~\ref{Table2}. The crystal field potential splits the $2J+1$ degenerate levels of the Pr$^{3+}$-ion into nine singlets with an overall splitting of 284~K.  It may be noted here that the ground state and the first excited state are separated by only 9~K. These two closely spaced singlet levels essentially form a quasi-doublet ground state, so that a magnetic ordering occurs in PrSi, supported by a magnetic exchange which is apparently above the threshold value.  
\begin{table*}
\caption{\label{Table2} Crystal field parameters and the corresponding energy levels obtained from the CEF calculation of the reciprocal susceptibility of PrSi.}
\begin{tabular}{|c|c|c|c|c|c|c|c|c|c|}
\hline 
\multicolumn{10}{|c|}{Crystal Field Parameters (K)}\tabularnewline
\hline 
\hline 
$B_2^0$ & $B_2^2$ & $B_4^0$ & $B_4^2$ & $B_4^4$ & $B_6^0$ & $B_6^2$ & $B_6^4$ & $B_6^6$ & $\lambda_i (i = x,y, z)$\tabularnewline
\hline 
\multirow{3}{*}{-3.22} & \multirow{3}{*}{0.81} & \multirow{3}{*}{-0.021} & \multirow{3}{*}{-0.072} & \multirow{3}{*}{-0.215} & \multirow{3}{*}{0.002} & \multirow{3}{*}{-0.008} & \multirow{3}{*}{0.013} & \multirow{3}{*}{0.002} & 17 mol/emu\tabularnewline
\cline{10-10} 
 &  &  &  &  &  &  &  &  & 13~mol/emu\tabularnewline
\cline{10-10} 
 &  &  &  &  &  &  &  &  & 10~mol/emu\tabularnewline
\hline
\end{tabular}

\begin{tabular}{|c|c|c|c|c|c|c|c|c|}
\hline 
\multicolumn{9}{|c|}{~~~~~~Energy levels (K)~~~~~~}\tabularnewline
\hline 
\hline 
~~~~0~~~~ & ~~~~9~~~~ & ~~~~29~~~~ & ~~~~67~~~~ & ~~~~97~~~~ & ~~~~148~~~~ & ~~~~161~~~~ &  ~~~~184~~~~  & ~~~~284~~~~  \tabularnewline
\hline 
\end{tabular}
\end{table*}

In the orthorhombic system, Bowden~\textit{et al,}~\cite{Bowden} and Shohata~\cite{Shohata}, have shown that the paramagnetic Curie-temperature $\theta_{\rm a}$, $\theta_{\rm b}$ and $\theta_{\rm c}$ and the $B_2^0$ and $B_2^2$  crystal field parameters can be expressed by the following expressions:

\begin{equation}
\begin{split}
\label{eqn4}
\theta_{\rm a} = \theta_{\rm \lambda} + (2J - 1)(2J + 3) \frac{(B_2^0 + B_2^2)}{10 k_{\rm B}}, \\
\theta_{\rm b} =  \theta_{\rm \lambda} + (2J - 1)(2J + 3) \frac{B_2^0}{5 k_{\rm B}}, \\
\theta_{\rm c} = \theta_{\rm \lambda} + (2J - 1)(2J + 3) \frac{(B_2^0 - B_2^2)}{10 k_{\rm B}}, 
\end{split}
\end{equation}
where, $\theta_{\rm \lambda}$ is the paramagnetic Curie temperature due to the molecular field and is given by the following expression,
\begin{equation}
\label{eqn5}
\theta_{\rm \lambda} = \lambda g_{\rm J}^2 \mu_{\rm B}^2 \frac {(J (J + 1))}{3 k_{\rm B}}.
\end{equation}

From the experimental values of the paramagnetic Curie-Weiss temperatures, for the three principal directions, the $B_2^0$ and $B_2^2$  parameters were estimated to be -3.22~K and 0.81~K respectively which  are in good agreement with the parameters obtained from our crystal field calculation.

\begin{figure}[!]
\includegraphics[width=0.5\textwidth]{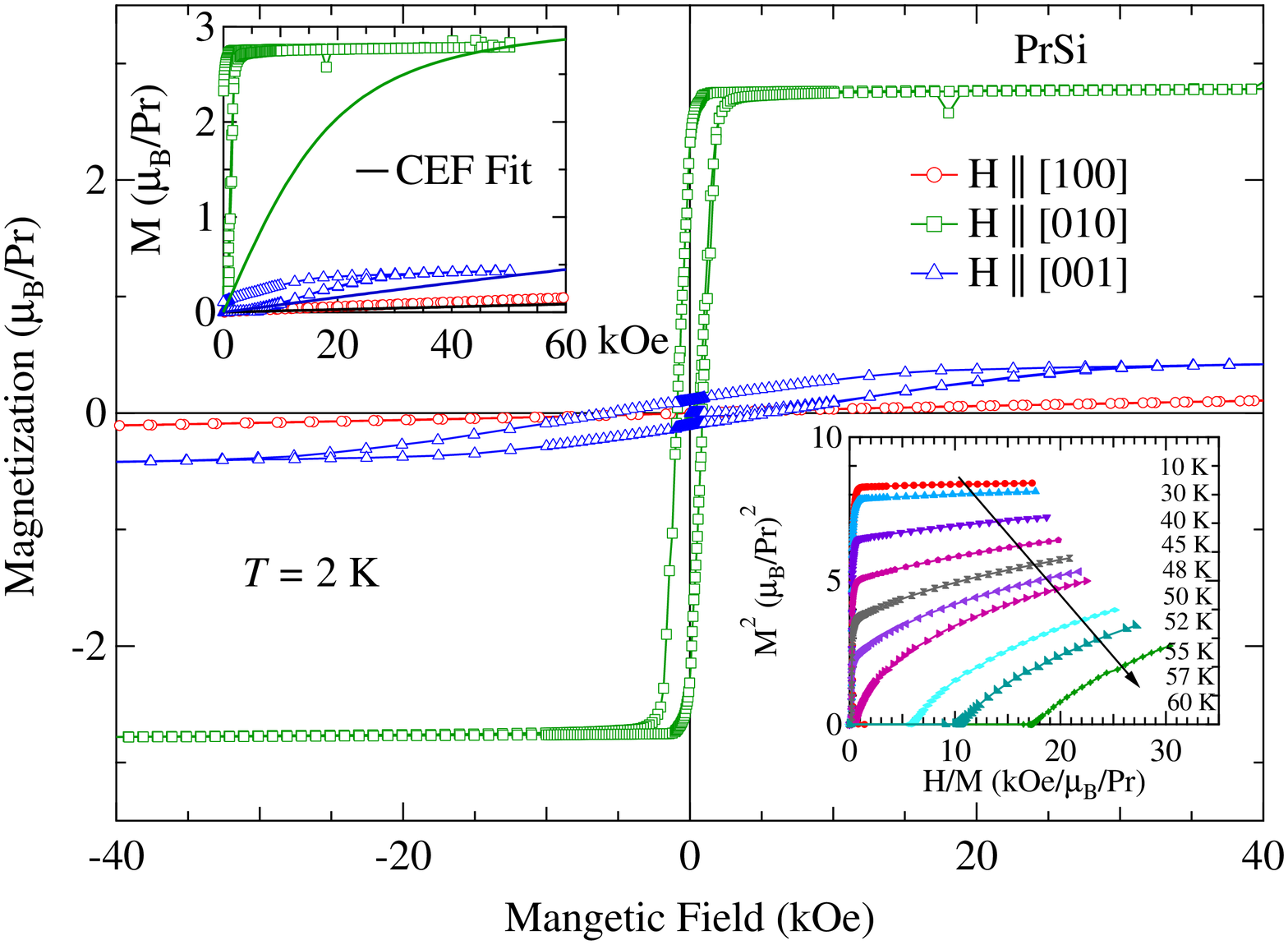}
\caption{\label{Fig5} Isothermal magnetization of PrSi along the three principal crystallographic directions. The upper inset shows calculated magnetization based on the crystal field model and the lower inset shows the Arrot plot.}
\end{figure}

The anisotropic magnetic behavior of PrSi was further investigated by measuring the isothermal magnetization at $T=2$~K well below ferromagnetic ordering temperature $T_{\rm C} = 52$~K and the data are shown in Fig.~\ref{Fig5}. The magnetization measurements were performed in both increasing and decreasing applied fields. The magnetization curves clearly corroborate the magnetization versus temperature curves shown in Fig.~\ref{Fig3}(a) with large anisotropy between the three principal directions. Hysteretic behavior is observed along all the three directions, confirming the ferromagnetic ground state in PrSi. For field parallel to [010], the magnetization increases very rapidly and saturates at fields greater than 3~kOe, thus confirming [010] as the easy axis of magnetization. The saturation moment is 2.88~$\mu_{\rm B}$/Pr which is very close to the saturation value of Pr$^{3+}$ free ion given by $g_{\rm J}J = 3.2$~$\mu_{\rm B}$/Pr. The magnetization along [100] and [001] directions reaches only 0.13~$\mu_{\rm B}$/Pr and 0.43~$\mu_{\rm B}$/Pr respectively indicating they lie in the hard plane of magnetization. The lower inset of Fig.~\ref{Fig5} shows the Arrott plots constructed in the temperature range from 10 to 60~K. The Arrot plots show an `S' shaped curvature which indicates a first order ferromagnetic transition~\cite{Banerjee}. Similar type of behavior was observed in the PrGe single crystals~\cite{Pranab_PrGe}.

We have calculated the magnetization curve based on the CEF model with the following Hamiltonian:
\begin{equation}
\label{eqn6}
\mathcal{H} = \mathcal{H_{\rm CEF}} - g_J \mu_{\rm B} J_{\rm i} (H + \lambda_{\rm i} M_{\rm i}), 
\end{equation}
where $\mathcal{H_{\rm CEF}}$ is given by Eq.~\ref{eqn2}, the second term is the Zeeman term and the third is the molecular field term. The magnetization $M_i$ is given by the following expression,

\begin{equation}
\label{eqn7}
M_i = g_J \mu_{\rm B}\sum_n \vert\langle n\vert J_i \vert n\rangle\vert \frac{{\rm exp}(-\beta E_n)}{Z},~~~~(i = x, y, z)
\end{equation}
By diagonalizing the total CEF Hamiltonian the energy eigenvalues $E_n$ and the eigenfunction $\vert n\rangle$ are obtained. The solid lines in the upper inset of Fig~\ref{Fig5} show the calculated magnetization curves based on the estimated crystal field parameters,   listed in Table~\ref{Table2}. The magnetocrystalline anisotropy is qualitatively explained by the set of crystal field parameters though the calculated magnetization curves do not match well with the corresponding experimental plots.

The magnetization value of a singlet ground state ferromagnet at 0~K can be estimated from the energy separation between the ground and the first excited singlet states $(\Delta)$ and the ferromagnetic Curie temperature $T_{\rm C}$ from the following expression~\cite{Yoshiuchi, Fazekas},

\begin{equation}
\label{eqn8}
m_{\rm T = 0~K} = 4 g_J \mu_{\rm B} \sqrt{1- tanh^2 \left(\frac{\Delta}{2 k_{\rm B} T_{\rm C}}\right)}.
\end{equation}
Using  $\Delta = 9$~K and  $T_{\rm C}=52$~K, results in a theoretical value of 3.18~$\mu_{\rm B}$/Pr which is in close agreement with our experimental saturation value of 2.88~$\mu_{\rm B}$/Pr.  The small discrepancy in the measured magnetization value may be attributed to the fact that the measurement was done at $T=2$~K.  Thus our crystal field calculation assuming an orthorhombic crystal potential is justified.  

\subsection{Heat Capacity}
We have measured the specific heat of PrSi in the temperature range from 1.8 to 200~K. In order to estimate the magnetic entropy of PrSi, we have also measured the specific heat of the non-magnetic reference compound 
\begin{figure}[!]
\includegraphics[width=0.4\textwidth]{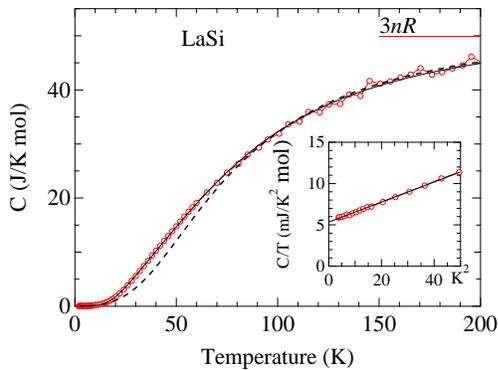}
\caption{\label{Fig6} Temperature dependence of specific heat capacity of LaSi.  The dashed line and the continuous line are fits to Debye model and Debye+Einstein model (see text for details). The inset shows the low temperature part of $C/T$ versus $T^{2}$ plot.  The solid line in the inset is the fit to $C/T = \gamma + \beta T^2$ expression.}
\end{figure}
LaSi. The specific heat of LaSi from 1.8 to 200~K is shown in the Fig.~\ref{Fig6} and it is typical of a non-magnetic reference compound. The inset in Fig.~\ref{Fig6} shows the low temperature part of $C/T~{\rm vs}~T^2$ plot.  The Sommerfeld coefficient $\gamma$ and the lattice contribution $\beta$ were obtained by fitting the data to $C/T = \gamma + \beta T^2$.  A best fit to the data was obtained for $\gamma = 5.341$~mJ/K$^2 \cdot$mol and $\beta = 1.2127~\times 10^{-4}$~mJ/K$^{4} \cdot$mol. The Debye temperature of LaSi can be estimated from the  value of $\beta$ through the expression $\Theta_{\rm D} = (1943.7~\times~n/\beta)^{1/3}$, where $n$ corresponds to the number of atoms (for LaSi, $n = 2$). The $\Theta_D$ value thus obtained is 317~K. However, this value of $\Theta_{\rm D}$ does not result in a good fit to the experimental data using the simple Debye model as shown by the dotted lines in Fig.~\ref{Fig6}. This suggests that the higher energy optical modes should also be considered in LaSi and hence we fitted the LaSi heat capacity data to the following expression:

\begin{equation}
\label{eqn9}
C_{\rm p} = \gamma T +\left[ m C_{\rm Debye}(T) + (1 - m) C_{\rm Einstein}(T)\right],
\end{equation}  
where the first term represents the electronic contribution and the second term represents the phononic contribution which includes the Debye plus the Einstein terms.  

The Debye and Einstein  terms are given by the following expressions, 
\begin{equation}
\label{eqn10}
C_{\rm Debye} = 9 n R \left(\frac{T}{\Theta_{\rm D}}\right)^3 \int_0^{\Theta_{\rm D}/T} \frac{x^4 e^x}{(e^x -1)^2} dx,
\end{equation}
and
\begin{equation}
\label{eqn11}
C_{\rm Einstein} = 3 n R \frac{y^2 e^y}{(e^y - 1)^2} ,
\end{equation}

where $x = \Theta_{\rm D}/T$ and $y = \Theta_{\rm E}/T$.  Equation~\ref{eqn9} results in a best fit to the experimental data of LaSi with 74\% of the weight to Debye term with $\Theta_{\rm D} = 373$~K and the remaining 26\% to the Einstein mode with Einstein temperature $\Theta_{\rm E} = 120$~K as shown by the solid line. At 200~K, the experimental heat capacity reaches a value of about 45~J/K~mol, which is close to the Dulong-Petit limiting value of $3nR~=~ 49.884$~J/K~mol.  

\begin{figure}[!]
\includegraphics[width=0.40\textwidth]{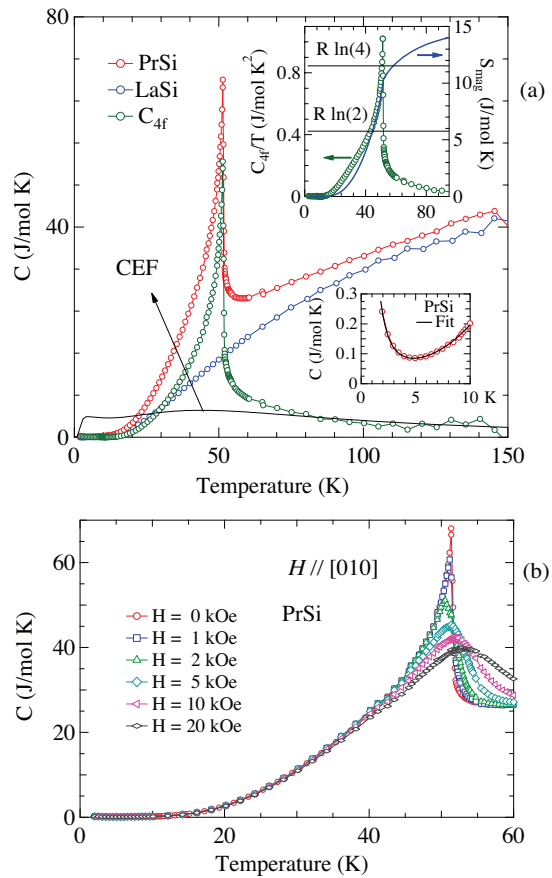}
\caption{\label{Fig7} (a) Heat capacity of PrSi and LaSi, and the magnetic part of heat capacity, the top inset shows $C_{\rm 4f}/T$ versus $T$ plot and the entropy. The lower inset in (a) shows the low temperature part of heat capacity and a fit to the nuclear Schottky and (b) Magnetic field dependence of heat capacity in PrSi.}
\end{figure}

The temperature dependence of specific heat capacity for single crystalline PrSi together with its non-magnetic analogue LaSi is shown in Fig.~\ref{Fig7}. As the temperature is decreased from 150~K, the heat capacity decreases smoothly and at 52~K a sudden jump is observed in the heat capacity which signals the bulk magnetic ordering of PrSi. The huge jump in the heat capacity at the magnetic transition amounts to 40~J/K~mol further confirming the first order nature of the magnetic transition as inferred earlier from the Arrot plots of the magnetization data.  Similar first order nature of the magnetic transition is also observed in the closely related PrGe single crystal~\cite{Pranab_PrGe}.   At  temperatures sufficiently below $T_{\rm C}$ for $T~\leq~5$~K,  the specific heat shows an upturn which is attributed to the nuclear Schottky contribution arising due to the interactions of the nuclear moments with $4f$-electrons. For $T~\leq~4$~K, the heat capacity has been fitted to following expression:
\begin{equation}
\label{eqn12}
C_{\rm p} = \gamma T + \beta T^3 + \left(\frac{C_{\rm N}}{T^2}\right),
\end{equation}  
where the first two terms are the usual electronic and lattice terms and the third term is the contribution from the nuclear Schottky heat capacity due to the splitting of the hyperfine levels.  Here we have ignored the magnon terms while fitting the heat capacity data as our fitting was performed for temperature much less than the magnon energy gap (about an order of magnitude less than the magnon gap estimated from the resistivity data).  The  fitting parameters thus obtained are $\gamma = 7.378$~mJ/K$^2 $mol, $\beta = 0.115~{\rm mJ/ K}^4$ mol and the nuclear Schottky term $C_{\rm N} = 850.2~{\rm mJ K}$/mol.   The coefficient $C_{\rm N}$ can be related to the magnetic moment of Pr moment $m_{\rm 4f}$ as

\begin{equation}
\label{eqn13}
C_{\rm N} = R  A_{\rm hf}^2 m_{\rm 4f}^2 \frac{I (I + 1)}{3 g_J^2},
\end{equation}

where $R$ is the gas constant, $A_{\rm hf}$ is the hyperfine coupling constant and $I$ is the nuclear spin.  For $^{141}$Pr, the nuclear spin $I = 5/2$ and the value of $A_{\rm hf}$  obtained from the literature is 0.052~K. Using these values in Eq.~\ref{eqn13}, the magnetic moment of Pr is obtained as 2.88~$\mu_{\rm B}$/Pr, which matches well with the saturation moment along the easy axis [010] as observed in the magnetization plot in Fig.~\ref{Fig5}.  This indicates that the sharp increase in the heat capacity at low temperature is not due to the presence of some magnetic impurities but it originates from the nuclear Schottky contribution arising from the hyperfine split nuclear levels.  The magnetic field dependence of the heat capacity of PrSi for $H~\parallel$~[010] is shown in Fig.~\ref{Fig7}(b).  There is a significant influence of the magnetic field on the specific heat.  As the magnetic field is increased the jump in the heat capacity decreases and the peak position shifts to higher temperature, as larger field aligns the ferromagnetic moments well above the zero field ordering temperature,  which is typically observed in ferromagnetic compounds.

We obtained the magnetic part $C_{\rm 4f}$ of the specific heat capacity of PrSi by subtracting the nuclear Schottky part and the phonon contribution obtained from the LaSi heat capacity data.  The magnetic entropy is obtained by the usual method of integrating the $C_{\rm 4f}/T$.  The entropy reaches $R~ln4$ just above the magnetic transition at 60~K which means that there are four $4f$ levels below that temperature which is consistent with our CEF calculations discussed earlier.  With further increase in temperature the entropy increases gradually due to the thermal population of the higher levels.  Just below the magnetic transition the entropy falls off more rapidly down to the value $R ln2$ and then it falls off gradually at lower temperatures. The Schottky contribution to the heat capacity is estimated using the following expression:

\begin{widetext}
\begin{equation}
\label{eqn14}
C_{Sch}=R\left[\frac{\displaystyle\sum\limits_{i}g_{i}e^{-E_{i}/T}\sum_{i}g_{i}E_{i}^{2}e^{-E_{i}/T}-\left[\sum_{i}g_{i}E_{i}e^{-E_{i}/T}\right]^{2}}{T^{2}\left[\displaystyle\sum\limits_{i}g_{i}e^{-E_{i}/T}\right]^{2}}\right],
\end{equation}
\end{widetext}
where $R$ is the gas constant, E$_{\mathrm{i}}$ is the CEF energy level in units of temperature and g$_{\mathrm{i}}$ the corresponding degeneracy. We have calculated the Schottky heat capacity based on the energy levels obtained from the CEF calculation of susceptibility data.  The solid line in Fig.~\ref{Fig7} shows the calculated heat capacity and it matches well with the magnetic part of the heat capacity at high temperature above the magnetic ordering.  This further justifies the validity of the crystal field parameters and the energy levels of nine singlets calculated from the magnetic susceptibility data.

\subsection{Magnetocaloric effect}

\begin{figure}
\includegraphics[width=0.45\textwidth]{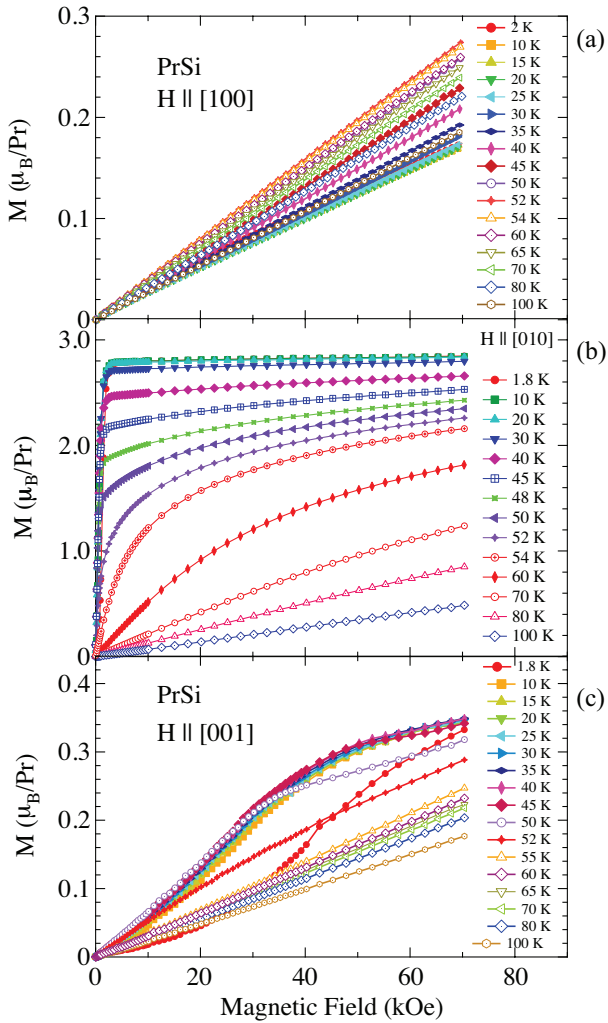}
\caption {Representative magnetization isothermals at various temperatures along (a) the $a$ axis ($H~\parallel~[100]$), (b) the $b$ axis ($H~\parallel~[010]$) and (c) the $c$ axis ($H~\parallel~[001]$). Only field increasing measurement processes is included for each temperature.}
\label{Fig8}
\end{figure}

The huge jump in the heat capacity at the magnetic transition and the Arrot plots together revealed a first order like magnetic transition.  We therefore decided to study the magnetocaloric properties of PrSi single crystal. For this purpose, we measured the magnetization of PrSi along the three principal crystallographic directions in the temperature range  $2$~K to $100$~K at selected  temperatures  and for field ranging from 0 to 70~kOe.  The  magnetization plots along the three directions are shown in Fig.~\ref{Fig8}.  Along the easy axis  [010] the variation of magnetization with temperature is relatively simple and along the expected lines and at temperature above the Curie temperature $T_{\rm C} = 52$~K the paramagnetic behaviour is observed. On the other hand along the hard axes, [100] and [001], the magnetization showed some anomalous behaviour.  Since there is a large magnetic anisotropy in PrSi, it is pertinent to calculate the individual  magnetic entropy change $(\Delta S_{\rm M})$ for the three directions respectively.  $\Delta S_{\rm M}$ in PrSi was calculated from $M~vs.~H$ isotherms using Maxwell's relation~\cite{gs}: $\Delta S_M (0 \rightarrow H) = \int_0^{H} {\frac{dM}{dT} dH}$.

\begin{figure}
\includegraphics[width=0.40\textwidth]{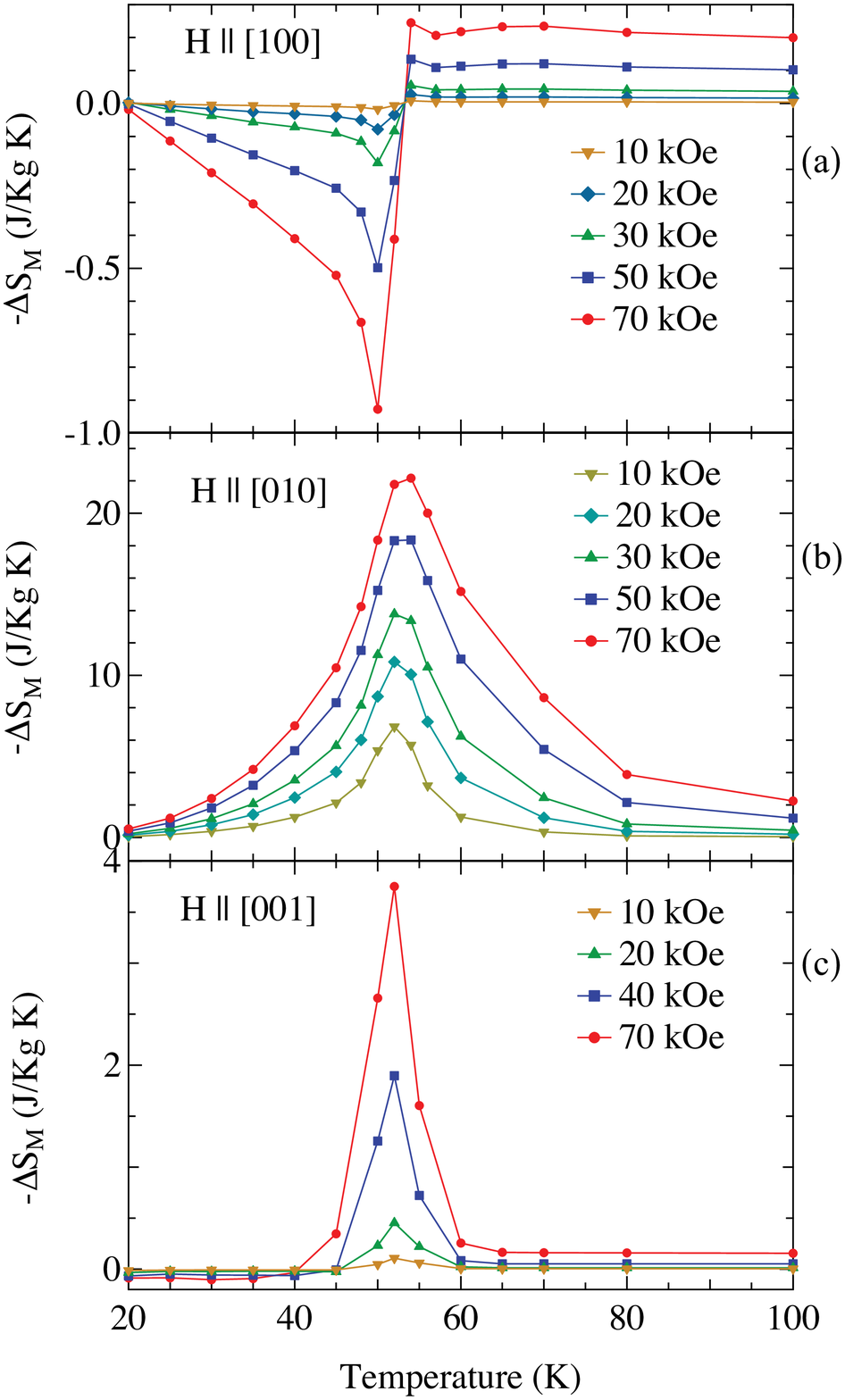}
\caption {(a), (b) and (c) represent the temperature variation of magnetic entropy change ($-\Delta S_M$, along the three principal crystallographic directions, $H~\parallel~[010]$, $H~\parallel~[001]$ and $H~\parallel~[100]$, respectively) around the magnetic transition region at various fields of PrSi. } 
\label{Fig9}
\end{figure}

\begin{figure}
\includegraphics[width=0.40\textwidth]{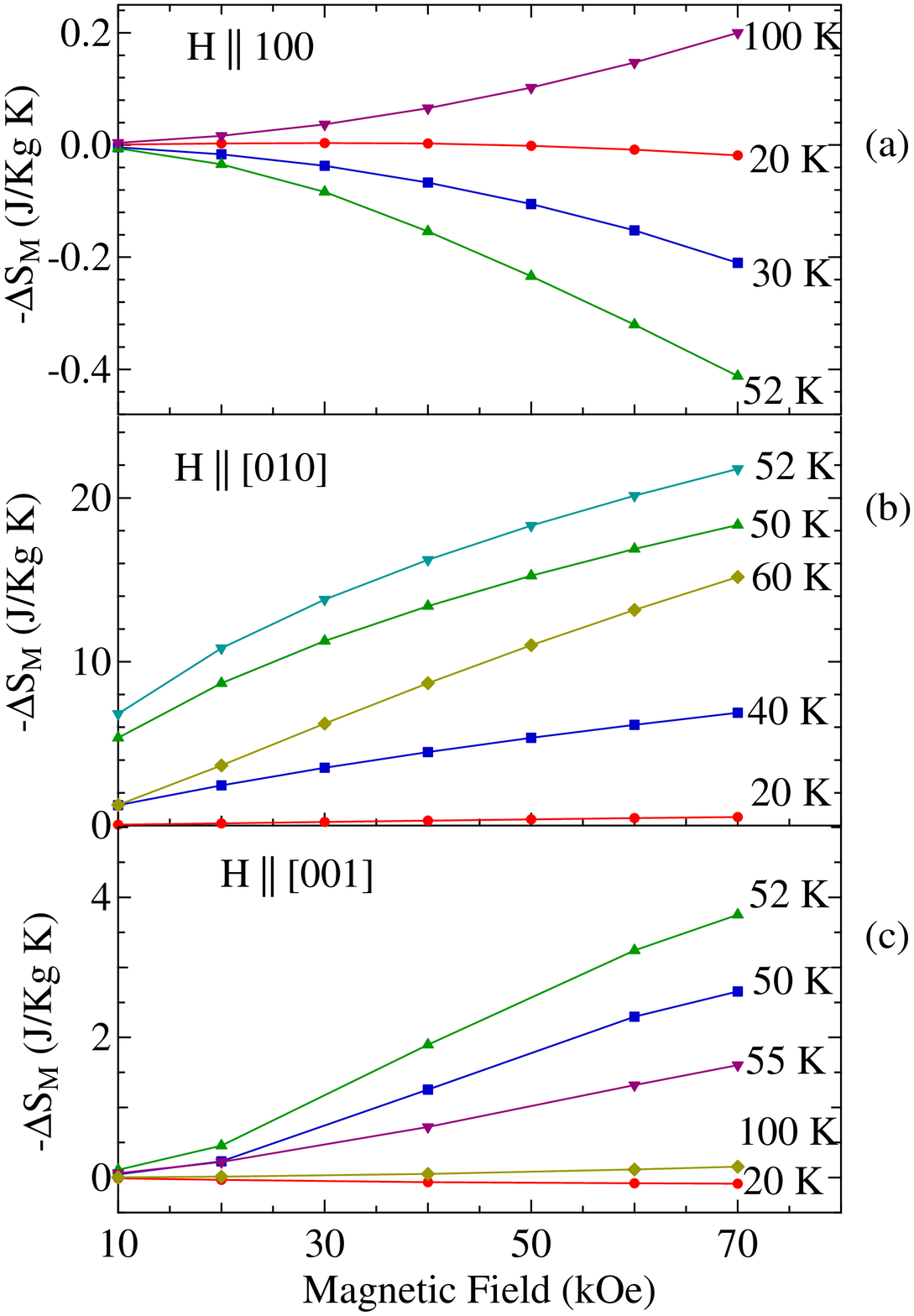}
\caption { (a), (b) and (c) represents the field variation of magnetic entropy change ($-\Delta S_M$, along the three principal crystallographic directions, $H~\parallel~$[010], $H~\parallel~[001]$ and $H~\parallel~ [100]$) around the magnetic transition region at different constant temperatures of PrSi.}
\label{Fig10}
\end{figure}

Temperature dependence of  the computed $-\Delta S_{\rm M}$ at selected values of the magnetic  field $H$, along the three principal directions,  respectively is shown in Fig.~\ref{Fig9}.  A low field is enough to induce complete saturation for $H~\parallel~[010]$ direction. Therefore, we observe giant negative magnetic entropy change  along this direction around $T_{\rm C}$. The maximum value of $-\Delta S_{\rm M}$ is found to be 22.2~J/kg~K  for $\Delta H$~=~70~kOe around $T_{\rm C}$ and it attains a value of 18.4~J/kg~K for a field change of 50~kOe. These values of $-\Delta S_{\rm M}$ are comparable to that of the well-known giant magnetocaloric compound Gd$_5$Si$_2$Ge$_2$~\cite{Pecharsky, Tishin}.  The $-\Delta S_{\rm M}$  value that we observe is larger than the value of PrSi polycrystalline sample~\cite{Strydom}.  Since the previous report was on polycrystalline sample they observed a lower MCE due to the large anisotropy.   Along $H~\parallel~[001]$ axis the magnetization is much smaller even for fields up to 70~kOe. Besides the $M-H$ data is non linear along this direction. Due to the lack of saturation here the value of $-\Delta S_{\rm M}$ is almost 10 times smaller than $H~\parallel~[010]$ axis. On the other hand, for  $H~\parallel~[100]$ direction the magnetic entropy is positive in the magnetically ordered state for fields as high as 70~kOe.    By comparing $-\Delta S_M$ for the $H~\parallel~[010]$, and $H~\parallel~ [100]$ directions, a giant anisotropy of $-\Delta S_{\rm M}$ is observed in PrSi single crystal.

\begin{figure}
\includegraphics[width=0.40\textwidth]{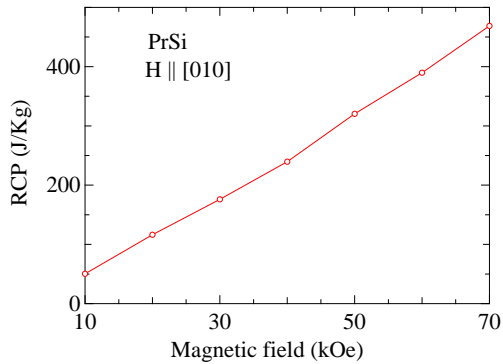}
\caption {Magnetic field variation of relative cooling power for PrSi along the [010] direction.} 
\label{Fig11}
\end{figure}

Figure~\ref{Fig10} shows the isothermal field variation of magnetic entropy change for a maximum applied field of 70~kOe.  For PrSi, clear non-linearity is observed in the $-\Delta S_M$ versus $H$ plot at low-$T$.  The general behavior of $-\Delta S_M$ versus $H$ for PrSi  is as follows: (i) linear temperature dependence well below the transition temperature (ii) around magnetic ordering temperature an $ H^{0.59}$ dependence (iii) quadratic temperature dependence in the paramagnetic region~\cite{franco}.

Relative cooling power (RCP) of a magnetic refrigerant is a measure of how much heat is transferred between the hot and cold sinks in one ideal refrigeration cycle and it is an important paramter as far as the magnetic  cooling is concerned~\cite{gs}.  For a magnetocaloric material the magnetic cooling efficiency is evaluated by using the relation,  $\rm{RCP} = |\Delta S_M^{max}| \delta T_{FWHM}$ \cite{ks}, where $\delta T_{FWHM}$ is the full width at half maximum of the -$\Delta S_M$ vs. $T$ curve and $ |\Delta S_M^{max}|$ is the maximum value of the entropy change. Isothermal field dependence of RCP for PrSi is shown in Fig.~\ref{Fig11}. RCP varies linearly with $H$.  RCP value is reasonably large, which is $\approx~470~$J/K  for a change of magnetic field $H$ to 70~kOe. It is very interesting to note here that usually, this type of giant magnetocaloric effect and large RCP is observed for samples possessing heavier rare-earth elements whose magnetic moments are large~\cite{ks, Tishin}.  In spite of the Pr atom's low saturation moment value (3.2~$\mu_{\rm B}$)  a large magnetocaloric effect is observed in PrSi. We are tempted to attribute the large MCE in PrSi to a magnetostructural transition at $T_{\rm C}$ such that the change in the configurational entropy adds to the magnetic entropy.

\section{Summary}

We have successfully grown a single crystal of  PrSi  and studied in detail its  anisotropic physical properties, including the magnetocaloric effect.  We conclude from  various measurements that PrSi  undergoes a ferromagnetic transition at $T_{\rm C} = 52$~K. Well defined Laue diffraction spots, together with a  low residual resistivity of the sample indicate the high quality of the grown single crystal. The magnetization measurements clearly reveal that the magnetic moments are oriented along the [010] direction.  Our results do not support the conclusion derived  from the previous neutron diffraction measurement where the magnetic moments were inferred to lie in the $ac$-plane.  Along  [100] and [001], the magnetization curve exhibit some anomalous behaviour by crossing each other at temperature near 10~K. This may be attributed to the presence of low lying crystal field levels and/or due to the complex nature of magnetic ordering in this compound. A detailed neutron diffraction measurement on a single crystal will throw more light on this.  Our crystal field calculations for the inverse susceptibility  revealed that the crystal field potential splits the $J=4$ degenerate ground state of Pr$^{3+}$-ion into nine singlets with an overall splitting energy of 284~K.  The ground  and the first excited state  separated by an energy difference of 9~K  essentially form  a quasi-doublet ground state and hence the magnetic ordering in this compound. From the energy of the first excited state and the ferromagnetic Curie temperature $T_{\rm C}$ the estimated magnetic moment at 0~K is found to be in agreement  with our experimental saturation moment measured at $T = 2$~K. The specific heat shows a huge jump at the ferromagnetic transition, which is usually observed when the transition is first order like.  From the specific heat jump and the Arrot plot, evidence for the first order nature of the transition  from the paramagnetic to ferromagnetic state has been obtained.  The low temperature heat capacity of PrSi shows an upturn which is well explained by means of the nuclear Schottky anomaly arising due to the hyperfine splitting of the nuclear levels of $^{141}$Pr.  The Pr $4f$ magnetic moment estimated from the coefficient of the nuclear Schottky heat capacity is in good agreement with our experimentally measured magnetic moment.  The magnetocaloric studies also revealed a large anisotropy along the three principal crystallographic directions, with a large relative cooling power for $H$ parallel to [010]. The giant magnetocaloric effect indicates the application potential of PrSi for the purpose of magnetic refrigeration in the low temperature range.

\end{document}